\documentclass[preprint,aps]{revtex4}
\usepackage{graphicx}
\usepackage{dcolumn}
\usepackage{bm}

\begin{document}
\title{{\bf Hard Sphere Dynamics for Normal and Granular Fluids}}
\author{James W. Dufty and Aparna Baskaran}
\affiliation{Department of Physics, University of Florida,
Gainesville, FL 32611}

\date{\today }

\begin{abstract}
A fluid of $N$ smooth, hard spheres is considered as a model for
normal (elastic collisions) and granular (inelastic collisions)
fluids. The potential energy is discontinuous for hard spheres so
the pairwise forces are singular and the usual forms of Newtonian
and Hamiltonian mechanics do not apply. Nevertheless, particle
trajectories in the $N$ particle phase space are well defined and
the generators for these trajectories can be identified. The first
part of this presentation is a review of the generators for the
dynamics of observables and probability densities. The new results
presented in the second part refer to applications of these
generators to the Liouville dynamics for granular fluids. A set of
eigenvalues and eigenfunctions of the generator for this Liouville
dynamics is identified in a special "stationary representation".
This provides a class of exact solutions to the Liouville equation
that are closely related to hydrodynamics for granular fluids.

\end{abstract}
\maketitle

\section{Introduction}

The properties of simple atomic fluids are typically well-described by
classical mechanics of point particles with phenomenological pairwise
additive potentials. These potentials are chosen to capture the quantitative
features of the actual quantum mechanical description of electronic
interactions between pairs of nuclei: strongly repulsive effects at short
distances and weak attraction at large distances. Except at low temperatures
the short range repulsion dominates and the classical potentials behave
qualitatively as $(\sigma /r)^{-n}$ where $r$ is the distance between a
pair, $\sigma $ is a characteristic force range, and $n$ is a large integer
(e.g., $n=12$ for the Lennard-Jones potential). The primary effect of the
phenomenological potentials is therefore one of excluded volume, dominating
all thermodynamic, structural, and transport properties. This suggests a
final idealization to consider, the hard sphere limit $n\rightarrow \infty $%
, where the potential is unbounded for $r<\sigma $ and zero for $r>\sigma $.
The single parameter of the hard sphere fluid, $\sigma $, then can be chosen
to give a reasonable quantitative description of real fluids \cite{McQuarrie}%
.

The equilibrium statistical mechanics of hard spheres reduces to a
determination of configurations for non-overlapping spheres, a geometrical
problem. All thermodynamic and structural properties of the hard sphere
fluid can be understood simply as a limiting form for strongly repulsive
potentials. In contrast, the dynamical properties of hard spheres provide
new problems due to the singular forces and vanishing time scale for pair
collisions. Consequently, the description of trajectories in the $N$
particle phase space cannot be given directly by Newton's or Hamilton's
equations \cite{Ernst,Dom,Resibois,vanB,McL}. Nevertheless, it is clear that
such trajectories exist and are uniquely defined for the usual initial data.
All particles move freely until a given pair is at contact, then the momenta
of that pair is changed instantaneously according to some chosen collision
rule (different for elastic and inelastic collisions, as described in the
next section). Subsequently the particles continue to move freely until the
next pair at contact occurs, the collision rule is applied instantaneously
to that pair, and free streaming is again resumed. The resulting
trajectories are unique, deterministic, and reversible up to the initial
data. Some of the complexities of the crossover from strongly repulsive
continuous potentials to the hard sphere limit have been described recently 
\cite{Dufty1}.

In the next section a representation of the generators for hard sphere
dynamics is recalled \cite{Ernst,Dom,Resibois,vanB,McL}. Such hard sphere
generators have been applied with great success over the past thirty years
to explore many of the subtleties of transport phenomena in normal fluids
(e.g., nonanalytic density dependencies, algebraic time decay)\cite
{Resibois,McL}. More recently, interest in hard sphere fluids has been
revived as models for granular flows \cite{Brey,vanNoije}. Real granular
media consist of mesoscopic sized grains (e.g., sand, seeds, pharmaceutical
pills) which, when activated,  behave like complex fluids, i.e., fluids
whose constituent particles have non-trivial internal degrees of freedom. It
turns out that many of the qualitative and quantitative differences between
normal and granular fluids are captured by a model the hard sphere fluid
whose constituents undergo \textit{inelastic} or dissipative collisions. For
an overview of recent developments see the edited volumes of reference \cite
{Poschel1}, and the new text by Poschel \cite{Poschel2}.

Following the definitions of generators, attention is focused in section 3
on their application to the statistical mechanics of granular media. The
Liouville equation is described and shown to have no stationary solutions
(equilibrium) for an isolated system. Nevertheless, a change of variables to
accommodate the collisional ''cooling'' due to the inelastic collisions
leads to a representation supporting a stationary solution. Next, it is
shown that the existence of the stationary solution implies certain
solutions to the eigenvalue problem for the Liouville operator in this
representation. In this way a class of special solutions to the Liouville
equation is identified. It is noted that the eigenvalues are the same as
those for the exact macroscopic balance equations for mass, energy, and
momentum in the long wavelength limit. This suggests that the microscopic
excitations are the precursors of macroscopic hydrodynamics in a granular
fluid.

\section{Generators for Hard Sphere Dynamics}

The system of interest is a one component fluid of $N$ identical smooth hard
disks or spheres (mass $m$, diameter $\sigma $). The position and velocity
coordinates of the fluid particles will be denoted by $\left\{\mathbf{q}%
_{i}, \mathbf{v}_{i}\right\}$. The state of the system at time $t$ is
completely characterized by the positions and velocities of all particles at
that time and is represented by a point $\Gamma _{t}\equiv \left\{ \mathbf{q}%
_{1}(t),\ldots ,\mathbf{q}_{N}(t),\mathbf{v}_{1}(t),\dots ,\mathbf{v}%
_{N}(t)\right\} $ in the associated $2dN$ dimensional phase space, where $%
d=2 $ for hard disks and $d=3$ for hard spheres. The dynamics consists of
free streaming (straight line motion along the direction of the velocity at
time $t$, until any pair of particles, say $i,j$, is in contact. At the
contact time the relative velocity $\mathbf{g}_{ij}=\mathbf{v}_{i}-\mathbf{v}%
_{j}$ of that pair changes instantaneously according to the collision rule 
\begin{equation}
\widetilde{\mathbf{g}}_{ij}=\mathbf{g}_{ij}-\left( 1+\alpha \right) \left( 
\widehat{\mbox{\boldmath $\sigma$}}\cdot \mathbf{g}_{ij}\right) \widehat{%
\mbox{\boldmath $\sigma$}}.  \label{2.1}
\end{equation}
Here $\widehat{\sigma}$ is a unit vector directed from the center of
particle $j$ to the center of particle $i$ through the point of contact. The
parameter $\alpha $ (the coefficient of normal restitution) is chosen \emph{%
a priori} in the range $0<\alpha \leq 1$ and remains fixed for a given
system. The value $\alpha =1$ corresponds to elastic, energy conserving
collisions, while $\alpha <1$ describes an inelastic collision with a
corresponding energy loss for the pair. However, the center of mass velocity
is unchanged so that the total mass and momentum of the pairs are conserved
for all values of $\alpha $. Subsequent to the change in relative velocity
for the pair $i,j$ the free streaming of all particles continues until
another pair is at contact, and the corresponding instantaneous change in
their relative velocities is performed. The sequence of free streaming and
binary collisions determines uniquely the positions and velocities of the
hard particles at time $t$ for given initial conditions.

\subsection{Statistical Mechanics}

\label{sec2}The statistical mechanics for a fluid of inelastic hard spheres
has been described elsewhere \cite{Brey,vanNoije}. It is comprised of the
dynamics just described, a state specified in terms of a probability density 
$\rho (\Gamma )$, and a set of observables denoted by $A(\Gamma )$. The
expectation value for an observable at time $t>0$ for a state $\rho (\Gamma )
$ given at $t=0$ is defined by 
\begin{equation}
\langle A(t);0\rangle \equiv \int d\Gamma \rho (\Gamma )A(\Gamma _{t}),%
\hspace{0.3in}\Gamma _{t}\equiv \left\{ \mathbf{q}_{1}(t),\ldots ,\mathbf{q}%
_{N}(t),\mathbf{v}_{1}(t),\dots ,\mathbf{v}_{N}(t)\right\}   \label{2.2}
\end{equation}
where $A(\Gamma ,t)=A(\Gamma _{t})$, $\Gamma =\Gamma _{t=0},$ and $\Gamma
_{t}\equiv \left\{ \mathbf{q}_{1}(t),\ldots ,\mathbf{q}_{N}(t),\mathbf{v}%
_{1}(t),\dots ,\mathbf{v}_{N}(t)\right\} $. The dynamics can be represented
in terms of the generator $L$ defined by 
\begin{equation}
\langle A(t);0\rangle =\int d\Gamma \,\rho (\Gamma )e^{tL}A(\Gamma ).
\label{2.3}
\end{equation}
For continuous potentials the generator is easily recognized from Hamilton's
equation as a Poisson bracket operation with the corresponding Hamiltonian.
However, its identification for the discontinuous hard sphere potential is
less direct.

There are two components to the generator, corresponding to the two steps of
free streaming and velocity changes at contact. The first part is the same
as for continuous potentials while the second part replaces the contribution
from the singular force by a ''binary collision operator'' $T(i,j)$ for each
pair $i,j$ 
\begin{equation}
L=\sum_{i=1}^{N}\mathbf{v}_{i}\cdot \mathbf{\nabla }_{i}+\frac{1}{2}%
\sum_{i=1}^{N}\sum_{j\neq i}^{N}T(i,j).  \label{2.4}
\end{equation}
The binary collision operator $T(i,j)$ for continuous potentials is
identified directly from the Poisson bracket of Hamilton's equations
\begin{equation}
T(i,j)\rightarrow \theta _{ij}=-m^{-1}\left( \nabla _{q_{i}}V(\left| \mathbf{%
q}_{i}-\mathbf{q}_{j}\right| )\right) \cdot \left( \nabla _{v_{i}}-\nabla
_{v_{j}}\right) .  \label{2.5}
\end{equation}
For hard spheres, the position variables are still continuous functions of
time but the momenta are piecewise constant (in the absence of external
forces) and discontinuous. Thus, a general phase function has the form
\begin{eqnarray}
A(\Gamma _{t}) &=&\Theta \left( t_{1}-t\right) A\left( \left\{ \mathbf{q}%
_{1}(t),\ldots ,\mathbf{q}_{N}(t),\mathbf{v}_{1}(0),\dots ,\mathbf{v}%
_{N}(0)\right\} \right)  \nonumber \\
&&+\sum_{p=1}\Theta \left( t-t_{p}\right) \Theta \left( t_{p+1}-t\right)
A\left( \left\{ \mathbf{q}_{1}(t),\ldots ,\mathbf{q}_{N}(t),\mathbf{v}%
_{1}(p),\dots ,\mathbf{v}_{N}(p)\right\} \right) .  \label{2.6}
\end{eqnarray}
Here $\left\{ t_{p}\right\} $ are the times for the colliding pairs and $%
\left\{ \mathbf{v}_{i}(p)\right\} $ are the velocities in the time interval
between collisions $p$ and $p+1$. The Heaviside theta functions identify the
time intervals between collisions. Direct differentiation of this form leads
to the identification of the binary collision operator for hard spheres
\begin{equation}
T(i,j)=\Theta (-\mathbf{g}_{ij}\cdot \widehat{\mathbf{q}}_{ij})|\mathbf{g}%
_{ij}\cdot \widehat{\mathbf{q}}_{ij}|\delta (q_{ij}-\sigma )(b_{ij}-1),
\label{2.7}
\end{equation}
where $\mathbf{q}_{ij}$ is the relative position vector of the two
particles, $\Theta $ is the Heaviside step function, and $b_{ij}$ is a
substitution operator
\begin{equation}
b_{ij}A(\mathbf{g}_{ij})=A(b_{ij}\mathbf{g}_{ij})=A(\widetilde{\mathbf{g}}%
_{ij}),  \label{2.7a}
\end{equation}
which changes the relative velocity $\mathbf{g}_{ij}$ into its scattered
value $\widetilde{\mathbf{g}}_{ij}$, given by Eq.\ (\ref{2.1}). The theta
function and delta function in (\ref{2.7}) assure that a collision takes
place, i.e. the pair is at contact and directed toward each other. Further
details of this approach can be found in Appendix A of reference \cite
{Lutsko} by Lutsko . Alternatively, the same result can be obtained through
a limiting procedure starting with a continuous potential whose slope
increases without bound.

An alternative equivalent representation of the dynamics is to transfer it
from the observable $A(\Gamma )$ to the state $\rho (\Gamma )$ by the
definition
\begin{equation}
\int d\Gamma \,\rho (\Gamma )e^{tL}A(\Gamma )\equiv \int d\Gamma \,\left(
e^{-t\overline{L}}\rho (\Gamma )\right) A(\Gamma ).  \label{2.8}
\end{equation}
The representation in terms of a dynamical state is referred to as Liouville
dynamics. Implicit in the analysis of the previous paragraph for hard
spheres is the restriction of the phase space to non-overlapping
configurations. This is assured when the generator $L$ is used in the
context of averages such as (\ref{2.3}) since all acceptable probability
densities $\rho (\Gamma )$ must include a singular factor excluding the
domain of any overlapping pair. However, the right side of (\ref{2.8}) no
longer has that restriction and consequently the generator for Liouville
dynamics is not the same as that for observables (as in the case of
continuous potentials). Instead, direct analysis of (\ref{2.8}) leads to the
result 
\begin{equation}
\overline{L}=\sum_{i=1}^{N}\mathbf{v}_{i}\cdot \mathbf{\nabla }_{i}-\frac{1}{%
2}\sum_{i=1}^{N}\sum_{j\neq i}^{N}\overline{T}(i,j),  \label{2.9}
\end{equation}
with the new binary collision operator
\begin{equation}
\overline{T}_{-}(i,j)=\delta (q_{ij}-\sigma )|\mathbf{g}_{ij}\cdot \widehat{%
\mathbf{q}}_{ij}|(\Theta (\mathbf{g}_{ij}\cdot \widehat{\mathbf{q}}%
_{ij})\alpha ^{-2}b_{ij}^{-1}-\Theta (-\mathbf{g}_{ij}\cdot \widehat{\mathbf{%
q}}_{ij})).  \label{2.10}
\end{equation}
Here $b_{ij}^{-1}$ is the inverse of the operator $b_{ij}$ in (\ref{2.7a}) 
\begin{equation}
b_{ij}^{-1}\mathbf{g}_{ij}=\mathbf{g}_{ij}-\frac{1+\alpha }{\alpha }\left( 
\widehat{\mbox{\boldmath $\sigma$}}\cdot \mathbf{g}_{ij}\right) \widehat{%
\mbox{\boldmath $\sigma$}}.  \label{2.11}
\end{equation}

In summary, the problems presented by the singular forces for a fluid of
hard spheres are resolved if Hamilton's equations for observables are
replaced by 
\begin{equation}
\left( \partial _{t}-L\right) A(\Gamma ,t)=0,  \label{2.12}
\end{equation}
and the Liouville equation for probability densities is replaced by
\begin{equation}
\left( \partial _{t}+\overline{L}\right) \rho (\Gamma ,t)=0,  \label{2.13}
\end{equation}
with the respective generators given by (\ref{2.4}) and (\ref{2.9}). An
important observation emphasized by Lutsko \cite{Lutsko} is that the form of
the generator $L$ and corresponding binary collision operator $T(i,j)$ does
not depend on the details of the collision rule defining the operator $%
b_{ij} $. In particular, the result applies for both elastic and inelastic
collisions. In contrast, the generator for Liouville dynamics is obtained by
a change of variables that introduces the Jacobian of the transformation
between the variables $\mathbf{g}_{ij}$ and $b_{ij}\mathbf{g}_{ij}$. Hence
it depends explicitly on the collision rule, or for granular media on the
restitution coefficient $\alpha $.

\section{Liouville Equation for Granular Fluids}

In the remainder of this presentation attention will be focused on the
granular fluids case $\alpha <1$, for an isolated system. In contrast to
normal fluids, there is no stationary solution to the Liouville equation for
an isolated system. This follows by calculating from it the average
''thermal'' speed of the particles $v(t),$ defined in terms of the average
kinetic energy 
\begin{equation}
v^{2}(t)\equiv \frac{4}{3mN}\left\langle \sum_{i=1}^{N}\frac{1}{2}m{v}%
_{i}^{2}(t);0\right\rangle .  \label{3.1}
\end{equation}
where $m$ is the mass, and the second equality defines the average . Using
either representation (\ref{2.12}) or (\ref{2.13}) the average speed is
found to be monotonically decreasing
\begin{equation}
\partial _{t}\ln v(t)=-\frac{1}{2}\zeta (t),  \label{3.2}
\end{equation}
where $\zeta (t)>0$ is the ``cooling'' rate due to inelastic collisions, 
\begin{equation}
\zeta \left( t\right) =(1-\alpha ^{2})\frac{N}{6v^{2}(t)}\int d\Gamma (%
\mathbf{g}_{12}\mathbf{\cdot }\widehat{\mathbf{q}}_{12})^{3}\Theta (\mathbf{g%
}_{12}\cdot \widehat{\mathbf{q}}_{12})\delta (q_{12}-\sigma )\rho (\Gamma
,t).  \label{3.3}
\end{equation}
This shows that there is no ''approach to equilibrium'' for a granular fluid
since there is no such stationary equilibrium state. Nevertheless, it is
expected (on both theoretical grounds and computer simulation results) that
there is a universal state that is approached for a wide range of initial
preparations.

This universal state, the homogeneous cooling solution (HCS), is special in
the sense that it is spatially homogeneous (translationally invariant) and
all of its time dependence occurs through the average speed $v(t).$ If there
are no other externally imposed energy scales, dimensional analysis requires
that the dependence on $v(t)$ occurs only through normalization and through
a scaling of the velocities, 
\begin{equation}
\rho _{hcs}=(\ell v_{hcs}\left( t\right) )^{-3N}\rho _{hcs}^{\ast }\left( \{%
\frac{\mathbf{q}_{ij}}{\ell },\frac{\mathbf{v}_{i}-\mathbf{u}}{v_{hcs}(t)}%
\}\right) .  \label{3.4}
\end{equation}
Here $v_{hcs}(t)$ is the thermal speed evaluated for the HCS, $\mathbf{u}$
is the constant average velocity of the system and $\ell $ is a constant
characteristic length (e.g., the mean free path). In this state, collisional
cooling due to inelastic collisions amounts only to a monotonic decrease in
all velocities. Use of (\ref{3.4}) in (\ref{3.3}) shows that the cooling
rate in the HCS has the time dependence
\begin{equation}
\zeta _{hcs}\left( t\right) =\frac{v_{hcs}(t)}{\ell }\zeta _{hcs}^{\ast },
\label{3.4a}
\end{equation}
where $\zeta _{hcs}^{\ast }$ is a dimensionless constant. Then (\ref{3.2})
can be solved directly to get the explicit time dependence of $v_{hcs}(t)$%
\begin{equation}
v_{hcs}(t)=v_{hcs}(0)\left( 1+\frac{v_{hcs}(0)}{2\ell }\zeta _{hcs}^{\ast
}t\right) ^{-1}\rightarrow \frac{2\ell }{\zeta _{hcs}^{\ast }t}.
\label{3.4b}
\end{equation}
Also given is the limiting behavior at long times, showing that the scaling
becomes independent of the initial conditions.

\subsection{Stationary Representation}

More generally, collisional cooling affects both the form of the
distribution function as well as scaling the velocities. It is useful to
account for the latter by making a time dependent change of variables to
obtain a dimensionless form of the Liouville equation
\begin{equation}
\mathbf{q}_{i}^{\ast }=\frac{\mathbf{q}_{i}}{\ell },\hspace{0.3in},\mathbf{V}%
_{i}^{\ast }=\frac{\mathbf{v}_{i}-\mathbf{u}}{\omega (t)}\hspace{0.3in}ds=%
\frac{\omega (t)}{\ell }dt,  \label{3.5}
\end{equation}
\begin{equation}
\rho ^{\ast }\left( \{\mathbf{q}_{i}^{\ast },\mathbf{V}_{i}^{\ast
}\},s\right) =(\ell \omega \left( t\right) )^{3N}\rho (\Gamma ,t).
\label{3.6}
\end{equation}
Here $\omega (t)$ is a characteristic velocity whose form is to be chosen
below. In terms of these variables the dimensionless Liouville equation
becomes
\begin{equation}
\partial _{s}\rho ^{\ast }-\ell \frac{\stackrel{\cdot }{\omega }(t)}{\omega
^{2}(t)}\sum_{i=1}^{N}\nabla _{\mathbf{V}_{i}^{\ast }}\cdot (\mathbf{V}%
_{i}^{\ast }\rho ^{\ast })+\overline{L}^{\ast }\rho ^{\ast }=0,  \label{3.7}
\end{equation}
with the time independent dimensionless generator 
\begin{equation}
\overline{L}^{\ast }=\frac{\ell }{\omega (t)}\overline{L}_{-}=\left[ 
\overline{L}_{-}\right] _{\{\mathbf{v}_{i}=\mathbf{V}_{i}^{\ast }\}}
\label{3.8}
\end{equation}
Next, $\omega (t)$ is chosen to \ make \ the coefficients of the Liouville
equation independent of $s$%
\begin{equation}
-\ell \frac{\stackrel{\cdot }{\omega }(t)}{\omega ^{2}(t)}=\omega ^{\ast
}\equiv 1,  \label{3.9}
\end{equation}
where $\omega ^{\ast }$ is an arbitrary dimensionless constant taken here to
be unity. The solution to this equation is
\begin{equation}
\omega (t)=\omega (0)\left( 1+\frac{\omega (0)}{\ell }t\right) ^{-1},
\label{3.10}
\end{equation}
and consequently the new time scale is
\begin{equation}
s=\ln \left( 1+\frac{\omega (0)}{\ell }t\right)  \label{3.11}
\end{equation}
The final form for the dimensionless Liouville equation is now 
\begin{equation}
\partial _{s}\rho ^{\ast }+\overline{\mathcal{L}}^{\ast }\rho ^{\ast }=0,
\label{3.12}
\end{equation}
with the modified generator
\begin{equation}
\overline{\mathcal{L}}^{\ast }\rho ^{\ast }=\overline{L}^{\ast }\rho ^{\ast
}+\sum_{i=1}^{N}\nabla _{\mathbf{V}_{i}^{\ast }}\cdot (\mathbf{V}_{i}^{\ast
}\rho ^{\ast })  \label{3.13}
\end{equation}

This time dependent change of variables was first suggested by Lutsko \cite
{Lutsko2}, motivated by the objective of molecular dynamics simulation of
the HCS. Direct simulation in the original variables is difficult at long
times since the velocities monotonically scale toward zero. In contrast, the
corresponding solution to (\ref{3.12}) is now a stationary solution\ $\rho
_{0}^{\ast }$ given by
\begin{equation}
\overline{\mathcal{L}}^{\ast }\rho _{0}^{\ast }=0.  \label{3.14}
\end{equation}
Consequently, the corresponding thermal speed for the new variables
approaches a constant
\begin{eqnarray}
v^{\ast 2}(t) &=&\frac{v^{2}(t)}{\omega ^{2}(t)}\left\langle \sum_{i=1}^{N}\frac{%
1}{2}m\mathbf{v}_{i}^{2}(t);0\right\rangle =\frac{2}{3N}\int d\Gamma ^{\ast
}\rho ^{\ast }\left( \Gamma ^{\ast },s\right) \sum_{i=1}^{N}\mathbf{v}%
_{i}^{\ast 2}  \nonumber \\
&\rightarrow &\frac{2}{3N}\int d\Gamma ^{\ast }\rho _{0}^{\ast }\left(
\Gamma ^{\ast }\right) \sum_{i=1}^{N}\mathbf{v}_{i}^{\ast 2}\equiv v^{\ast
2}(\infty )  \label{3.14a}
\end{eqnarray}
Since both $v_{hcs}(t)$ and $\omega (t)$ decay as $t^{-1}$ at long times it
follows that the dimensionless cooling rate $\zeta _{hcs}^{\ast }$ is simply
related to $v^{\ast }(\infty )$%
\begin{equation}
\zeta _{hcs}^{\ast }=\frac{2}{v^{\ast }\left( \infty \right)}.  \label{3.15}
\end{equation}
Since (\ref{3.12}) supports a stationary state it is referred to here as the
stationary representation of the Liouville equation.

\subsection{Special Solutions to the Liouville Equation}

The existence of the stationary state $\rho _{0}^{\ast }$ of \ the generator 
$\overline{\mathcal{L}}^{\ast }$ given by (\ref{3.14}) has some interesting
consequences. The stationary solution is assumed known. In the original
variables it satisfies the equations
\begin{equation}
\rho _{0}(\Gamma ,t)=(\ell \omega \left( t\right) )^{-3N}\rho _{s}^{\ast
}\left( \{\mathbf{q}_{i}^{\ast },\frac{\mathbf{v}_{i}-\mathbf{u}}{\omega (t)}%
\}\right)  \label{3.16}
\end{equation}
\begin{equation}
\overline{L}\rho _{0}+\frac{\omega (t)}{\ell }\sum_{i=1}^{N}\nabla _{\mathbf{%
v}_{i}}\cdot (\left( \mathbf{v}_{i}-\mathbf{u}_{i}\right) \rho _{0})=0.
\label{3.17}
\end{equation}
Differentiating (\ref{3.17}) with respect to $\omega \left( t\right) $ at
constant $\{\mathbf{q}_{i}^{\ast },\mathbf{v}_{i}\}$ gives
\begin{equation}
\overline{L}\partial _{\omega }\rho _{0}+\frac{\omega (t)}{\ell }%
\sum_{i=1}^{N}\nabla _{\mathbf{v}_{i}}\cdot (\left( \mathbf{v}_{i}-\mathbf{u}%
_{i}\right) \partial _{\omega }\rho _{0})=-\ell ^{-1}\sum_{i=1}^{N}\nabla _{%
\mathbf{v}_{i}}\cdot (\left( \mathbf{v}_{i}-\mathbf{u}\right) \rho _{0}).
\label{3.18}
\end{equation}
But it also follows from (\ref{3.16}) that 
\begin{equation}
\partial _{\omega }\rho _{0}=-\omega ^{-1}\sum_{i=1}^{N}\nabla _{\mathbf{v}%
_{i}}\cdot (\left( \mathbf{v}_{i}-\mathbf{u}_{i}\right) \rho _{0}).
\label{3.19}
\end{equation}
Substitution of (\ref{3.19}) into (\ref{3.18}) and transformation back to
dimensionless variables gives the result 
\begin{equation}
\overline{\mathcal{L}}^{\ast }\Psi ^{(1)}=\Psi ^{(1)},\hspace{0.3in}\Psi
^{(1)}=\sum_{i=1}^{N}\nabla _{\mathbf{V}_{i}^{\ast }}\cdot (\mathbf{V}%
_{i}^{\ast }\rho _{0}^{\ast }).  \label{3.20}
\end{equation}
Thus, the stationary state implies directly one solution to the eigenvalue
problem for $\overline{\mathcal{L}}^{\ast }$.

Repeating this analysis, but differentiating with respect to the total
number of particles $N$ and the flow velocity $\mathbf{u}$ gives the
additional eigenvalues and eigenvectors
\begin{equation}
\overline{\mathcal{L}}^{\ast }\Psi ^{(2)}=0,\hspace{0.3in}\Psi
^{(2)}=\partial _{N}\rho _{0}^{\ast }  \label{3.21}
\end{equation}
\begin{equation}
\overline{\mathcal{L}}^{\ast }\mathbf{\Psi }^{(3)}=-\mathbf{\Psi }^{(3)},%
\hspace{0.3in}\mathbf{\Psi }^{(3)}=\sum_{i=1}^{N}\nabla _{\mathbf{V}%
_{i}^{\ast }}\rho _{0}^{\ast }.  \label{3.22}
\end{equation}
The eigenvalue $-1$ is three fold degenerate since the equation holds for
each component of the vector $\mathbf{\Psi }^{(3)}.$

These eigenvalues and eigenvectors allow construction of a special class of
exact solutions to the Liouville equation (\ref{3.12}) 
\begin{eqnarray}
\rho ^{\ast }\left( \Gamma ^{\ast },s\right) &=&e^{-\overline{\mathcal{L}}%
^{\ast }s}\left( \rho _{0}^{\ast }\left( \Gamma ^{\ast }\right) +c_{1}\Psi
^{(1)}\left( \Gamma ^{\ast }\right) +c_{2}\Psi ^{(2)}\left( \Gamma ^{\ast
}\right) +\mathbf{c}_{3}\cdot \mathbf{\Psi }^{(3)}\left( \Gamma ^{\ast
}\right) \right)  \nonumber \\
&=&\rho _{0}^{\ast }\left( \Gamma ^{\ast }\right) +c_{1}\Psi ^{(1)}\left(
\Gamma ^{\ast }\right) e^{-s}+c_{2}\Psi ^{(2)}\left( \Gamma ^{\ast }\right) +%
\mathbf{c}_{3}\cdot \mathbf{\Psi }^{(3)}\left( \Gamma ^{\ast }\right) e^{s}
\label{3.23}
\end{eqnarray}
where $\left\{ c_{i}\right\} $ are arbitrary constants. Note that
normalization is preserved since the integral over phase space of all
eigenfunctions vanish. In terms of the original variables this class of
solutions can also be written
\begin{equation}
\rho \left( \Gamma ,s\right) =\rho _{0}\left( \Gamma \right)
+\sum_{n=1}^{5}e^{-\lambda _{n}s}c_{n}\partial _{y_{n}}\rho _{0}\left(
\Gamma \right)  \label{3.24}
\end{equation}
with 
\begin{equation}
\lambda _{n}\longleftrightarrow \left( 1,0,-1,-1,-1\right) ,\hspace{0.3in}%
y_{n}\longleftrightarrow \left( \omega ,N,\mathbf{u}\right) .  \label{3.25}
\end{equation}

\subsection{Relationship to Hydrodynamics}

To provide some interpretation to the dynamical effects represented by these
eigenfunctions and eigenvalues it is useful to consider the exact
macroscopic balance equations for density $n$, momentum density $mn\mathbf{u}$%
, and energy density $\frac{3}{2}nk_{B}T$. These follow directly by using
either (\ref{2.12}) or (\ref{2.9}) to calculate the averages of the
corresponding local microscopic densities, with the results 
\begin{equation}
D_{t}n+n\nabla \cdot \mathbf{u}=0,  \label{3.26}
\end{equation}
\begin{equation}
D_{t}u_{i}+(mn)^{-1}\partial _{i}p+(mn)^{-1}\partial _{j}P_{ij}=0,
\label{3.27}
\end{equation}
\begin{equation}
\left( D_{t}+\zeta \right) T+\frac{2}{3n}p\nabla \cdot \mathbf{u}+\frac{2}{3n%
}\left( P_{ij}\partial _{j}u_{i}+\nabla \cdot \mathbf{q}\right) =0,
\label{3.28}
\end{equation}
where $D_{t}=\partial _{t}+\mathbf{u}\cdot \nabla $ is the material
derivative, $T(\mathbf{r},t)$ is the temperature, $\mathbf{u}(\mathbf{r},t)$
is the flow velocity, $\mathbf{q}$ is the heat flux, and $P_{ij}$ is
irreversible part of the momentum flux. Linearize these equations about the
HCS solution ($n_{hcs}=$ constant, $T_{hcs}(t)=mv_{hcs}^{2}(t)/2,$ $\mathbf{u%
}_{hcs}=\mathbf{0}$), and introduce the dimensionless variables
\begin{equation}
\delta n^{\ast }=\frac{\delta n}{n_{hcs}},\hspace{0.3in}\delta T^{\ast }=%
\frac{\delta T}{T_{hcs}(t)},\hspace{0.3in}\delta \mathbf{u}^{\ast }=\frac{%
\delta \mathbf{u}}{v_{hcs}\left( t\right) }.  \label{3.29}
\end{equation}
Then in the long wavelength limit, neglecting all spatial gradients, these
equations become
\begin{equation}
\left( \partial _{s}+0\right) \delta n^{\ast }\approx 0.  \label{3.30}
\end{equation}
\begin{equation}
\left( \partial _{s}+1\right) \left( \delta T^{\ast }+\frac{\partial \ln
\zeta _{hcs}}{\partial \ln n_{hcs}}\delta n^{\ast }\right) \approx 0.
\label{3.31}
\end{equation}
\begin{equation}
\left( \partial _{s}-1\right) \delta U_{i}^{\ast }\approx 0.  \label{3.32}
\end{equation}
Therefore, it is seen that excitations obtained for the Liouville equation
are the same dynamics as the long wavelength limit for the macroscopic
balance equations.

\qquad More generally, it is expected that spatially inhomogeneous states
for a granular gas can be described by hydrodynamics. Such hydrodynamic
equations follow from the above macroscopic balance equations with
constitutive relations expressing the heat flux and momentum flux in terms
of spatial gradients of $n$, $\mathbf{u}$, and $T$. The resulting linearized
equations define the hydrodynamic modes for a granular system, which
necessarily behave as (\ref{3.30}) - (\ref{3.31}) at long wavelengths. This
shows that the eigenfunctions and eigenvalues of the Liouville equation
found here can be interpreted as the long wavelength microscopic precursors
of macroscopic hydrodynamics. This point is illustrated in more detail
elsewhere \cite{Baskaran}.

\section{Conclusion}

The objectives of this brief presentation have been two fold. The first was
an introduction to the means to describe the dynamics of a hard sphere
fluid, in spite of the singular forces and failure of standard Newtonian and
Hamiltonian formalisms. In fact, this is a special case of a class of
problems involving piecewise continuous or discontinuous potentials (e.g.,
the finite step potential, hard spheres with attractive square well)\cite{Dom}. In
each case the effect of the singular force can be replaced by an associated
binary collision operator to define a generator for the dynamics.

The second objective was to report some new results for the Liouville
dynamics of inelastic hard spheres, an idealized model for granular fluids.
The inelasticity implies a monotonic loss of energy for an isolated system.
However, by a change of variables to accommodate the average decrease in
each particle's speed the Liouville equation is given a representation that
supports a stationary state. The existence of this stationary state implies
certain properties of the associated generator for the dynamics. For systems
with elastic collisions this would imply certain invariants. For inelastic
collisions the eigenfunctions identified here are no longer invariants, but
instead have a simple dynamics directly related to the collisional cooling
rate. This can be seen by choosing as the scaling function $\omega
(t)\rightarrow v_{hcs}(t)$. In this case the eigenvalues (\ref{3.25}) become 
$\lambda _{n}\longleftrightarrow \left( \frac{1}{2}\zeta _{hcs}^{\ast },0,-%
\frac{1}{2}\zeta _{hcs}^{\ast },-\frac{1}{2}\zeta _{hcs}^{\ast },-\frac{1}{2}%
\zeta _{hcs}^{\ast }\right) $. It was then noted that the eigenvalues of the
Liouville equation are the same as those of the macroscopic balance
equations for average number density, energy density, and flow velocity in
the long wavelength limit. Hence they are long wavelength hydrodynamic
modes, whenever a closed set of hydrodynamic equations apply. This has been
exploited recently to define hydrodynamic response functions for a granular
fluid and to derive corresponding exact Green-Kubo expressions for the
transport coefficients \cite{Baskaran}. This identification of eigenvalues and eigenvectors 
of the Liouville operator is a generalization of a similar analysis at the level of the 
Boltzmann kinetic theory \cite{DBModes}. The connection to hydrodynamics has been established
clearly in that context.

\begin{acknowledgments}
This research was supported in part by Department of Energy Grant
DE-FG02ER54677.
\end{acknowledgments}


\begin{thebibliography}{99}
\bibitem{McQuarrie}  D. McQuarrie. 1973. Statistical Mechanics.
HarperCollins, New York.

\bibitem{Ernst}  M. Ernst, J. Dorfman, W. Hoegy, and J. van Leeuwen. 1969. Physica 
\textbf{45}, 127; J. Sengers, M. Ernst, and D. Gillespie. 1972. J. Chem.
Phys. \textbf{56}, 5583.

\bibitem{Dom}  J. Domaradzki. 1977. Physica \textbf{86A}, 169.

\bibitem{Resibois}  P. Resibois and M. De Leener. 1977. Classical Kinetic
Theory of Fluids. John Wiley, New York.

\bibitem{vanB}  H. van Beijeren and M. Ernst. 1979. J. Stat. Phys. \textbf{21},
125.

\bibitem{McL}  J. A. McLennan. 1989. Introduction to Nonequilibrium
Statistical Mechanics. Prentice-Hall, New Jersey.

\bibitem{Dufty1}  J. Dufty. 2002. Molec. Phys. \textbf{100}, 2331; J. Dufty
and M. Ernst. 2004. Molec. Phys. In press.

\bibitem{Brey}  J. J. Brey, J. W. Dufty, and A. Santos. 1997. J. Stat. Phys. 
\textbf{87}, 1051.

\bibitem{vanNoije}  T. P. C. van Noije and M. H. Ernst. 2001. \emph{In} Granular
Gases. T. Poschel and S. Luding Eds. Springer, NY.

\bibitem{Poschel1}  Granular Gases. 2001. T. Poschel and S. Luding Eds. Springer, NY; 
Granular Gases Dynamics. 2003. T. Poschel and N.
Brilliantov Eds. Springer, NY.

\bibitem{Poschel2} Kinetic Theory of
Granular Gases. N. Brilliantov and T. Poschel. 2004. Oxford, NY.

\bibitem{Lutsko}  J. Lutsko. 2004. J. Chem. Phys. \textbf{120}, 6325.

\bibitem{Lutsko2}  J. Lutsko. 2001. Phys. Rev. E. \textbf{63}, 061211.

\bibitem{Baskaran}  A. Baskaran, J. Dufty, and J. Brey. Linear
Response and Hydrodynamics for a Granular Fluid. Unpublished.

\bibitem{DBModes} J. Dufty and J.J. Brey. 2003. Phys. Rev. E \textbf{68}, 030302;
J. Dufty, J.J. Brey and M.J. Ruiz-Montero. 2003. \emph{In} Granular Gases Dynamics. 2003. T. Poschel and N.
Brilliantov Eds. Springer, NY.
\end{thebibliography}
\end{document}